# Impact of Non-Hermiticity and Nonlinear Interactions on Disordered-Induced Localized Modes.

BHUPESH KUMAR,[1,3] AND PATRICK SEBBAH[1,2,*]

[1]*Department of Physics, The Jack and Pearl Resnick Institute for Advanced Technology, Bar-Ilan University, Ramat-Gan, 5290002 Israel*
[2]*Institut Langevin, ESPCI ParisTech CNRS UMR7587, 1 rue Jussieu, 75238 Paris Cedex 05, France*
[3]*bhupeshg2@gmail.com*
[*]*patrick.sebbah@biu.ac.il*

**Abstract:** If disorder-induced Anderson localized states have been observed experimentally in optics, their study remains challenging leaving a number of open questions unsolved. Among them, the impact on Anderson localization of non-Hermiticity, optical gain and loss, and more generally, nonlinearities has been the subject of numerous theoretical debates, without yet any conclusive experimental demonstration. Indeed, in systems where localized modes have reasonable spatial extension to be observed and investigated, their mutual interaction and coupling to the sample boundaries make it extremely difficult to isolate them spectrally and investigate them alone. Recently, we successfully exhibited localized lasing modes individually in an active disordered medium, using pump-shaping optimization technique. However, a one-to-one identification of the lasing modes with the eigenmodes of the passive system was not possible, as the impact of non-Hermiticity and nonlinear gain on these localized states was unknown. Here, we apply the pump-shaping method to fully control the non-Hermiticity of an active scattering medium. Direct imaging of the light distribution within the random laser allows us to demonstrate unequivocally that the localized lasing modes are indeed the modes of the passive system. This opens the way to investigate the robustness of localized states in the presence of nonlinear gain and nonlinear modal interactions. We show that, surprisingly, gain saturation and mode competition for gain does not affect the spatial distribution of the modes.

## 1. Introduction

Real eigenvalues and orthogonal eigenfunctions are, in general, hallmarks of Hermitian systems. But when a vibrating system starts to couple to its surroundings or when a non-uniform distribution of gain or loss is introduced, the modes become complex-valued and their eigenfunctions are no longer orthogonal. Being ubiquitous in nature, non-Hermitian Hamiltonians and complex potentials have attracted considerable attention recently [1]. In particular, non-Hermitian classical and quantum optics form a new paradigm where new degrees of freedom and unprecedented control over light propagation and dynamics are offered, leading to unexpected new features [2–4]. Open disordered media are an interesting example of non-conservative system, where the amount of non-Hermiticity can be tuned by varying the scattering strength of the medium. When scattering is sufficiently strong, transport is suppressed and modes are exponentially localized, away from the sample boundaries. This is the celebrated regime of Anderson localization [5,6], which has become an important framework for the understanding of transport in mesoscopic systems and multiple scattering of classical-wave [7,8]. As scattering strength is progressively reduced, modes become spatially extended and start to couple to their surroundings: Their lifetime shortens while inter-modal coupling becomes significant. Enhancement of the quantum-limited laser linewidth by the Petermann factor is an example of the effect of modal complexness and non-orthogonality [9,10].

Another way to introduce non-Hermiticity is by distributing gain or loss non-uniformly in the system. Complex potentials, where disorder is introduced in both real and imaginary parts, have recently attracted considerable attention as they question the well-established fundamentals of Anderson localization [11–15]. Non-Hermitian disorder has been mostly investigated theoretically in the tailor-made paradigm of coupled waveguides with alternating gain and loss. The interplay between Parity-Time (*PT*) symmetry and transverse disorder lead to the destruction of localization at the point of (*PT*)-symmetry breaking [16]. While both amplification and loss are found to enhance the exponential decay of the transmittance [17, 18], others predict a delocalization for small disorder strengths [11, 19]. Actually, purely imaginary random potentials [20] can also localize [21–23]. Most recently, non-Hermiticity was predicted to break down the one-parameter scaling theory of localization [13], which had firmly established that 1D and 2D systems are always localized [24]. In fact, the question of whether localization is enhanced or destroyed in active scattering media has been the matter of early intense debates in the framework of random lasers, where non-Hermiticity is naturally introduced by local loss and gain [17, 25, 26]. As pointed out in [25], above threshold, this question becomes even more complicated: When multimode random lasing occurs, mode repulsion [27], competition for gain and nonlinear modal cross talk dominates [28]. One may expect that gain saturation as well as mode competition will strongly affect disordered-induced modal confinement. This question has never been addressed experimentally. An interesting hallmark of non-Hermiticity in random lasers is their sensitivity to non-uniform pumping [29, 30], which offers new degrees of freedom to control their emission. It was shown that single mode operation at any desired selected emission wavelength might be achieved by shaping the gain distribution [31, 32]. Gain shaping has also been proposed to control the directivity of the laser emission [33]. More recently, this method was coupled to an imaging technique to exhibit localized lasing modes in a random microlaser [34]. The question remained however, if the lasing modes that were observed were the actual localized modes of the passive system or if the non-Hermiticity, both in modal loss and non-uniform gain distribution, had modified the nature of their spatial confinement.

In this paper, we investigate experimentally non-Hermitian localization in an active random system to address these questions. Using pump shaping technique, we experimentally investigate localized lasing modes of a 1D disorder photonic structure, where the scattering strength is adjusted to probe a regime where the localization length is smaller but comparable to the sample size and modes significantly overlap with each other. First we found that lasing modes maintain their spatial profile under local pumping, which can occur only if they are built on a particular resonance of the passive system. This establishes for the first time on experimental grounds the predicted one-to-one correspondence between lasing modes and quasimodes of the passive system in the localized regime [35, 36]. Only close to sample edges where leakage occurs and non-Hermiticity is stronger, are lasing modes found to be sensitive to the size of the pumping area, showing that they are a superposition of quasi-bound states of the passive system, in agreement with theoretical prediction [37]. With this result, we explore the impact of non-Hermiticity on passive localized states by testing the sensitivity of the corresponding lasing modes to e.g. non-uniform gain distribution. We show that when pump energy is increased, localized mode profiles remain unchanged. They are robust even when gain saturation is reached. Even more remarkable, when several lasing modes interact to compete for gain, their spatial profile is perfectly preserved, demonstrating that localized modes are insensitive to cross-saturation effect.

## 2. Results

Our study is based on our previously reported solid-state organic random laser [34], where Anderson localization of lasing modes was demonstrated. In this doped-polymer waveguide (DCM-doped PMMA thin layer), hundreds of parallel randomly-distributed air-grooves have been carved over a total length of 1 mm, using electron-beam lithography. Here, the geometric parameters (grooves depth and width) have been adjusted for the random laser to operate in a regime where modes extend over a significant part of the sample, this way favoring spatial and spectral mode overlap. The sample design and its fabrication are described in full details in [34]. The intensity profile of the pumping laser at 532 nm is controlled by a spatial light modulator working in reflection. This shaping of the pump intensity profile allows to select a spectral line in the multimode emission spectrum of the random laser. Details of the experimental setup and the iterative method used to optimize the pump profile which selects a particular mode are given in [34]. When the pump profile is optimized and single mode operation is enforced, the sample surface is image on a camera and the intensity profile of the selected lasing mode is retrieved. This is how the localized nature of the lasing modes was revealed in [34].

However, one cannot infer anything about the nature of the quasimodes of the passive scattering system (disordered system without gain). Indeed, it was shown [38] that lasing modes can be radically different from the modes of the passive system, calling on for an ab-initio laser theory beyond the semiclassical laser theory [39]. This is particularly true in weakly scattering media where modes are spatially extended. Because the system is open, the passive modes are determined by a non-Hermitian operator and do not form an orthogonal basis. The lasing modes and the modes of the passive cavity are therefore likely to be different. In strongly scattering media however, where modes are spatially confined, Hermiticity is recovered since states are localized away from the boundaries and are therefore less sensitive to the openness of the system. One can expect therefore that the localized modes will provide the lasing modes of the active medium. This is reminiscent of the Fox-Li modes [40] in a conventional laser, where lasing modes are the modes of the passive cavity, aside from a correction on their emission frequency (frequency pulling). This one-to-one correspondence between the lasing modes and the passive modes in a strongly scattering medium was predicted numerically in [35, 36]. Here, we demonstrate it for the first time experimentally, opening the perspective of exploring Anderson localized modes, via their corresponding lasing modes.

A direct comparison of passive and lasing modes via independent measurements is however difficult, since the modes of the passive system are short-lived and, therefore, not accessible to the experimental measure. Here, we resort to an indirect method, by testing the robustness of lasing modes to local pumping. Simply stated, if the profile of a lasing mode is insensitive to the spatial extension of the pump beam, this means that laser oscillations sustained by gain occur on a natural mode of the passive system. Therefore, both the lasing mode and the passive mode are identically spatially distributed, except for a possible pulling of the resonance frequency. However, if the mode changes with pump extension, this proves that the lasing mode is enforced by the pumping profile and would change with it. The effect of partial pumping on random lasers has been reviewed in [41].

It was shown numerically in [35, 42] that partial pumping does not alter the character of a well localized lasing mode, provided absorption is absent in the unpumped region. Localized modes of the passive random system act therefore as ordinary modes of a conventional laser cavity selected by the gain. This contrasts with lasing modes in the weakly scattering regime, for which lasing mode profile may significantly depend on the pump extension. It was shown in [30] that the more pumping deviates from uniform, the larger the number of passive quasimodes contributing to the lasing mode. The lasing mode is not identified with a single passive quasimode but with a combination of several quasimodes. This is a signature of the non-Hermitian nature of the passive system and the complexness of its natural solutions [37]. This contrasts with conventional optical cavities or strongly disordered systems in the Anderson regime, where quasimodes are preserved in the presence of non-uniform gain.

To test it, we first optimize the random laser to operate in singlemode regime at $\lambda = 600.20$ nm, corresponding to a lasing mode localized near the center of the sample. The optimized non-uniform pump profile which selects this mode is shown in Fig. 1b, together with the intensity spatial distribution of the selected lasing mode. The robustness of this mode to local pumping is tested by reducing the pumping area to a short (40 μm-long) pump stripe, applied at the location of the maximum intensity of the initial mode. At pump fluence of 32 $J/m^2$ (vs. 4.8 $J/m^2$ with full optimized pump profile), a lasing mode is excited, exactly at the same wavelength ($\lambda$ = 600.20 nm) as the initial mode (Fig. 1a and c). Remarkably enough, its spatial profile shown in Fig. 1d closely resembles the profile of the optimally-pumped lasing mode (Fig. 1b). The correlation coefficient which measures the similarity of the two profiles, is r= 82 %. We make the assumption that the residual difference results from uniform loss/absorption in the unpumped region. To support this interpretation, we add a small uniform pump over the whole sample to undo the loss/absorption. This way, we recover the exact profile of the initial mode, as shown in Fig. 1f. The correlation coefficient is now r=96%. That local pumping is sufficient to excite the initially optimally-pumped lasing mode, while preserving its spatial profile and emission frequency, confirms that the lasing mode is actually the mode of the passive system which has been selected by the gain, in agreement with theoretical predictions for well localized modes [35]. This is demonstrated here for a mode far from the sample boundaries, so that the system can be considered as Hermitian.

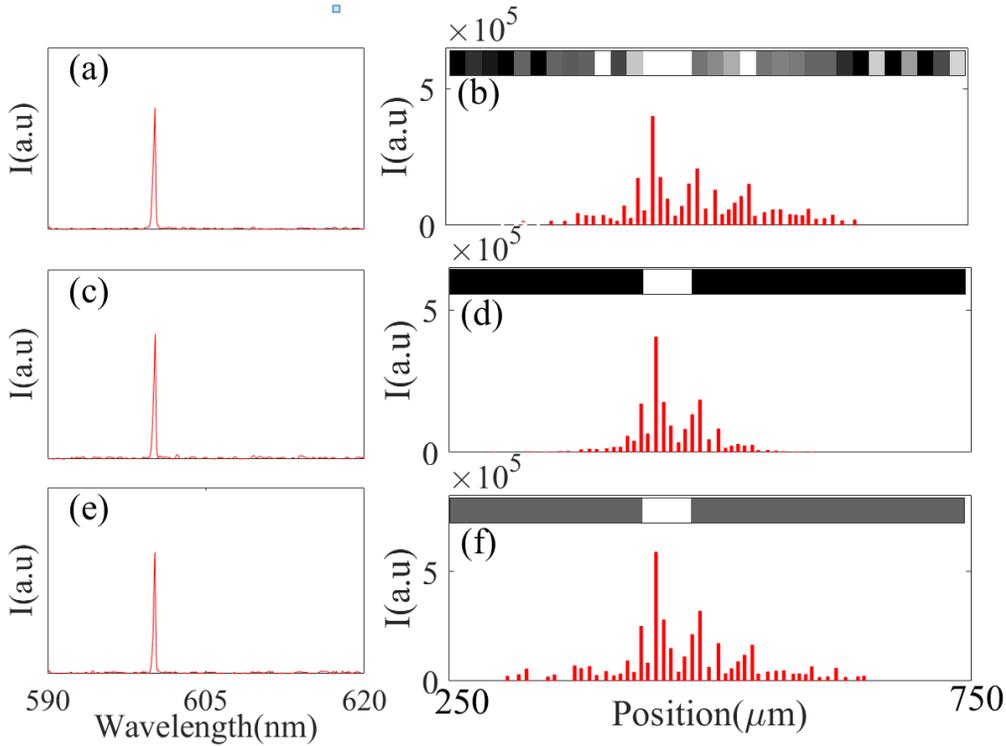

Fig. 1. **Local and optimized pumping of a mode localized away from the sample edges.** (a)&(b): Emission spectrum and spatial intensity profile of lasing mode @600.20 nm optimally-pumped near the middle of the sample (@pump fluence 4.8 $J/m^2$); (c)&(d) Emission spectrum and spatial intensity profile of lasing mode @600.20 nm locally-pumped (pump length is 40 $\mu$m, pump fluence 32 $J/m^2$). (e)&(f): Emission spectrum and spatial intensity profile of same locally-pumped lasing mode @600.20 nm after undoing loss by adding a small uniform pump. The corresponding pump intensity profile is shown on top of each intensity profile. Correlation coefficient between (b) and (d) is 80 %, while it rises to 96 % after undoing loss.

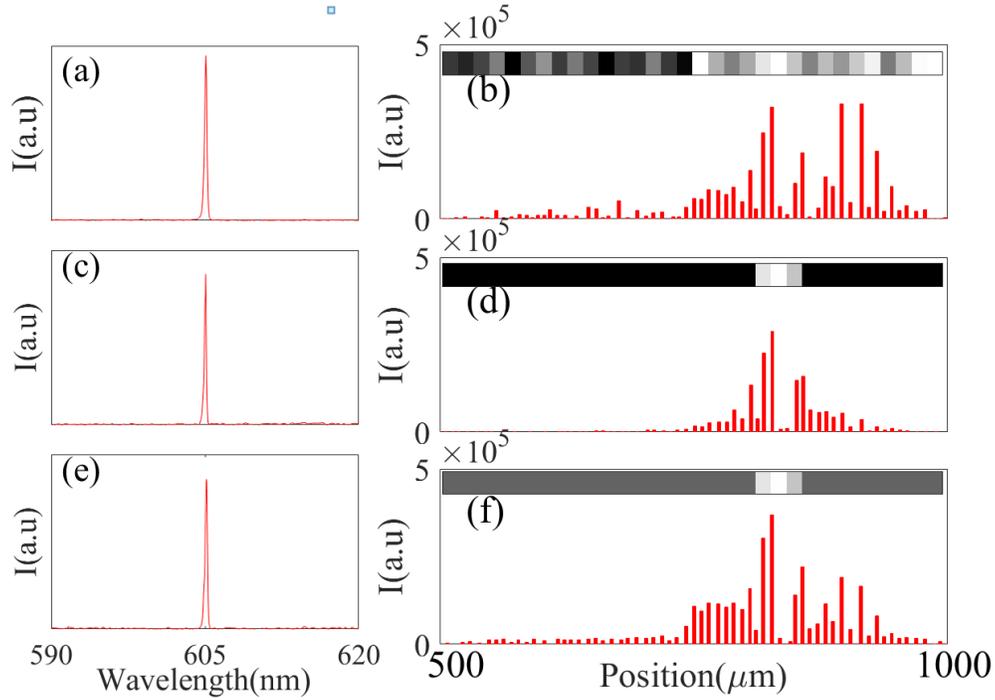

Fig. 2. **Local and optimized pumping of a mode localized near the sample edge.** (a)&(b): Emission spectrum and spatial intensity profile of lasing mode @605.21 nm optimally-pumped near the edge of the sample (@pump fluence 5.6 $J/m^2$); (c)&(d) Emission spectrum and spatial intensity profile of lasing mode @605.21 nm locally-pumped (pump length is 42 μm, pump fluence is 67.6 $J/m^2$). (e)&(f): Emission spectrum and spatial intensity profile of same locally-pumped lasing mode @605.21 nm after undoing loss by adding a small uniform pump. The corresponding pump intensity profile is shown on top of each intensity profile. Correlation coefficient between (b) and (d) is 56 %, while it rises to 90 % after undoing loss.

Non-Hermiticity and its impact on the nature of the mode selected by the gain can in turn be tested near the edge of the sample, where localized modes are leaky. Actually, it is relatively difficult experimentally to excite these modes, since they have high threshold. Nevertheless, we show here a measurable effect of non-Hermiticity for a mode near, but not at the sample boundary. The mode we select ($\lambda$ = 605.21 nm) is centered around x = 750 μm (total sample length 1000 μm), as shown in Fig. 2 b. The pumping area is subsequently reduced to a 42 μm-long local pump stripe near x=750 μm. A new lasing mode is forced at the same wavelength (Fig. 2a & c), but with a significantly different profile compare to that of the original mode (Fig. 2b & d). The correlation between the two profiles yields a correlation coefficient r= 56 %. By undoing uniform absorption/loss with an additional uniform pump, the correlation between the two lasing modes improves to r=90% (Fig. 2b & f). The residual discrepancy confirms however that the "edge" lasing mode is modified by local pumping. Therefore, the lasing mode cannot be identified with the mode of the passive system selected by the gain. The solutions of the laser equations can be significantly different from the quasimodes of the passive system, when they are short-lived, complex-valued, and not anymore orthogonal to each other. In conclusion of this section, the lasing modes perfectly reproduce the modes of the passive system, as long as they are selected sufficiently away from the sample boundaries. The localized nature of the quasimodes is therefore confirmed and their spatial intensity profile can be safely investigated by probing the corresponding lasing modes.

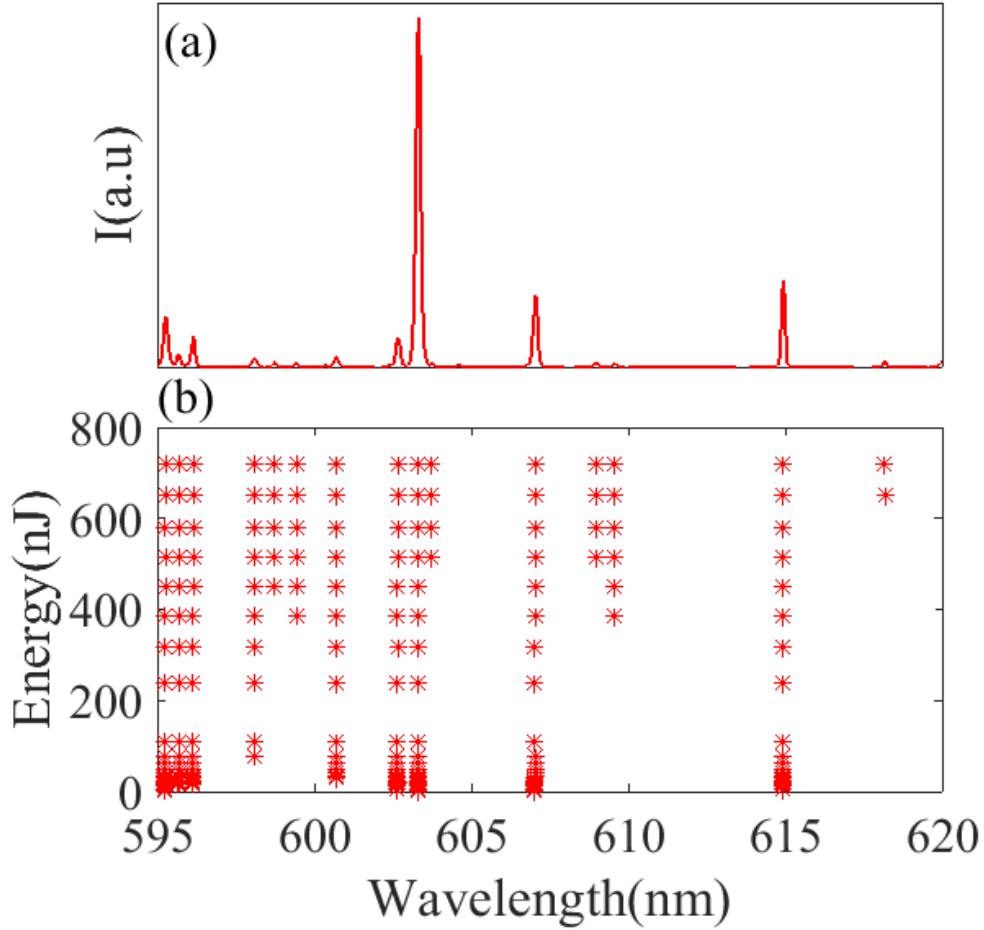

Fig. 3. **Spectral stability of multimode lasing under increasing uniform pumping.**
(a) Emission spectrum under uniform pumping, at pump energy 700 nJ, after averaging over 50 shots. (b) Spectral position of lasing modes plotted against pump energy.

With the ability to control it in single or multimode mode regime, our localized random laser offers a unique paradigm to explore this controversial question. Indeed, it brings together localized modes and nonlinear effects which actually have never been considered, such as nonlinear gain, gain saturation and mode competition. We first check the spectral stability of the laser emission at all pumping energy. The random laser is pumped uniformly over 500 $\mu$m X 50 $\mu$m. Lasing action is observed for pump energy above 18 nJ, when sharp discrete peaks appear in the emission spectrum (see Fig. 3a). The multimode emission spectrum is found to be stable at all pump energy. This is shown in Fig. 3b. The absence of energy pulling for all lasing wavelengths is a first indication that the corresponding localized quasimodes are robust to gain and do not suffer from nonlinear modal interactions in the multimode regime.

In a second experiment, the effect of nonlinear gain and gain saturation is examined on a single localized mode by increasing the pump energy and monitoring the mode profile. We first select a mode ($\lambda$ =603.10 nm) localized near the center of the sample. We check that this mode is preserved under local pumping and subsequently reduce the pump to a small 10 $\mu$m-long pump stripe applied near the mode center. Here we prefer a focused local pump over an optimized pump profile, to increase the energy density and make sure gain saturation can be reached without sample damage. The pump energy is increased step-wise from threshold level (14 nJ) to saturation (147 nJ) and beyond (200.3 nJ), as shown in Fig. 4. The lasing mode profile is recorded, as shown in Fig. 5. We observe that the mode profile is perfectly preserved, although its peak intensity has increased by more than three orders of magnitude and dye saturation has been reached. A quantitative comparison between the mode intensity profile measured at low pump energy and that measured at saturation gives a correlation coefficient of 97%. This is clear and direct evidence that the localized mode is insensitive to nonlinear gain, even in the presence of gain saturation. The robustness of localized modes to nonlinear gain has been tested and confirmed for all measurable localized modes of the system. Next, we investigate the impact of nonlinear modal crosstalk and gain competition on the localized nature of the modes. Using pump shaping, we first identify two localized lasing modes, mode 1 at $\lambda$ = 602.72 nm and mode 2 at $\lambda$ = 608.00 nm, which partially spatially overlap (33 %). Their spatial intensity profiles are shown in Fig. 6 together with the optimized pump profiles, P1 and P2 , which select each of them individually. In this experiment, we first pump mode 1 with pump profile P1 just above threshold (48 nJ). Next, we apply the second pump profile P2. This is simply done by superimposing both pattern on the SLM. While the P2 component of the pump energy is progressively increased, we monitor both the emission spectrum and the laser-field intensity distribution across the sample. The evolution of the emission spectrum shown in Fig. 7 dramatically illustrates the complex nonlinear competition and the mutual cross-saturation effects at stake between the two modes. When pump P2 is turned on, mode 1 takes this new resource to continue growing. This is made possible because the pump profiles P1 and P2 overlap spatially. Mode 2 is forced to remain below threshold, at a pump energy larger than its threshold level when pumped alone (55 nJ). However, when the total energy of the pump reaches 80 nJ, mode 2 competes then more efficiently for gain and starts lasing. Eventually, it saturates the gain provided to mode 1 and forces mode 1 to decrease. This partial mode-switching of mode 1 is an interesting signature of mode interaction [43]. In Supplement 1(section A), we demonstrate in another example complete switch-off of one mode as a result of cross gain saturation with a second mode. At a pump energy of 100 nJ, mode 1 takes control in turn and saturates mode 2. This no-win scenario of gain competition reproduces itself at 125 nJ, and 140 nJ. One may assume that such a strong gain competition would greatly modify the spatial profile of the modes and would affect localization. Actually, this is not the case at all. In Fig. 6f, we compare the intensity profile resulting from the simultaneous pumping at P1 and P2 to the profile of the modes selected individually (Fig. 6b & d). We find that this profile is precisely the weighed sum of the profiles of the two modes. The mode profile is therefore unaffected by cross-gain mode competition, a quite unexpected finding.

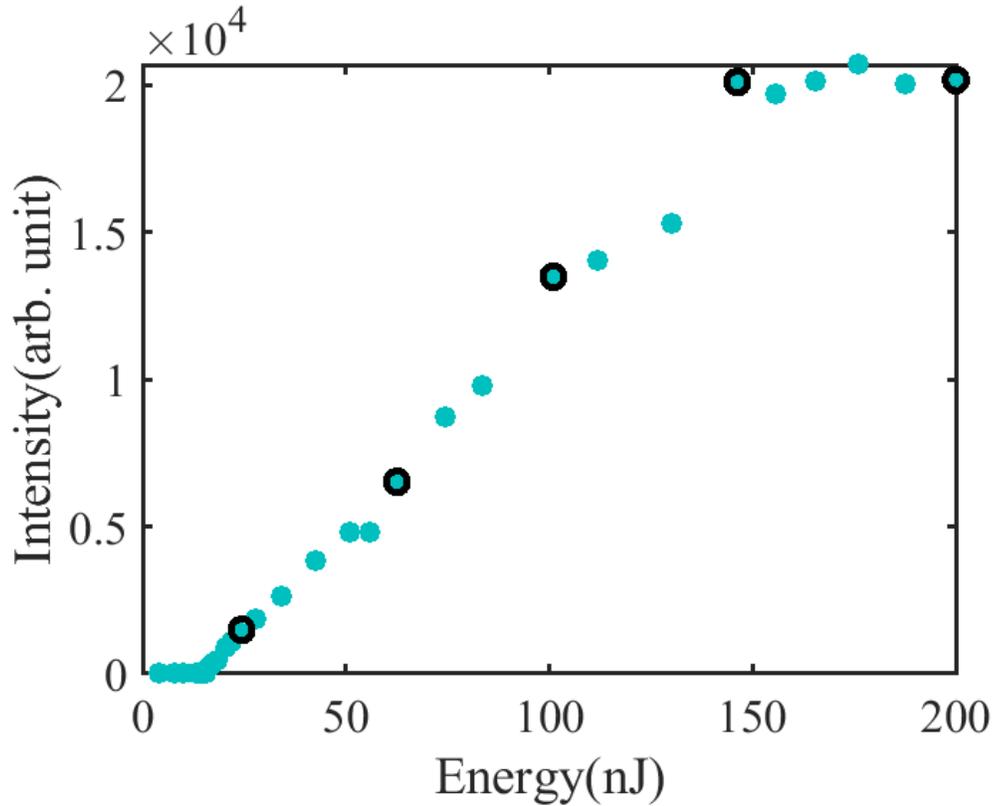

Fig. 4. **Gain saturation in singlemode operation.** Lasing characteristic for optimally-selected lasing mode @603.10 nm, operating in singlemode regime. Threshold is 14 nJ. Gain saturation is reached at 147 nJ. Black circles indicate the pump energies at which the mode intensity distribution is shown in Fig. 5.

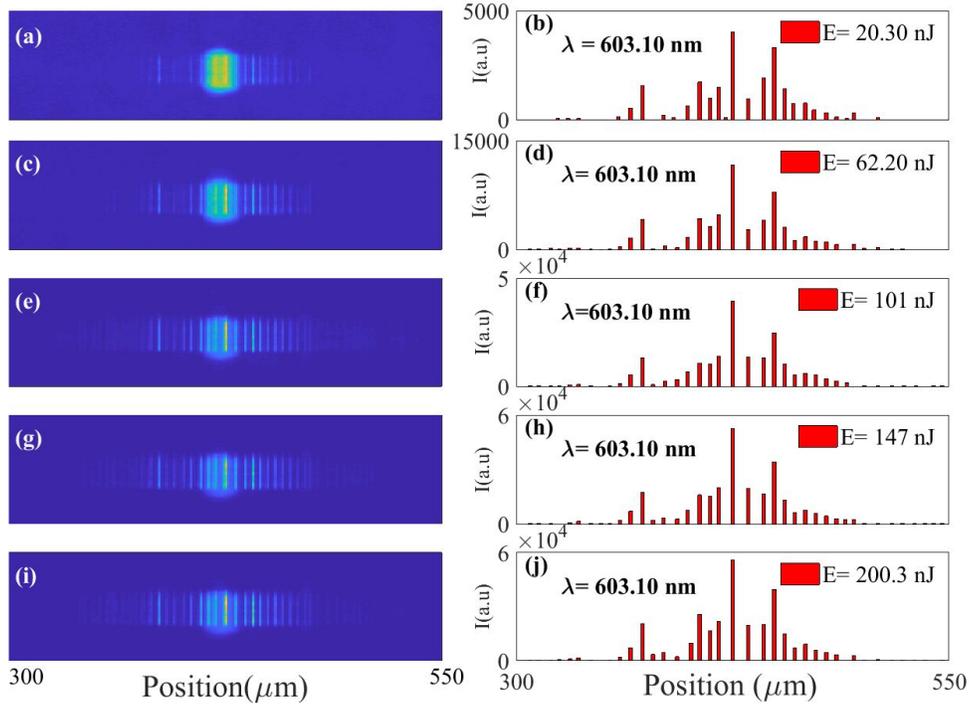

Fig. 5. **Localized mode robustness to gain nonlinearity.** Local pumping (10 μm-length) of mode @$\lambda$= 603.10 nm at increasing pump energy (a,b) 20 nJ, (c,d) 62.20 nJ, (e,f) 101 nJ, (g,h) 147 nJ, (i,j) 200.3 nJ. Left: Optical microscope image (10X) of the spatial intensity distribution showing both fluorescence near the pump area and lasing light scattered by grooves. Right: Corresponding lasing intensity at the grooves (bar plots).

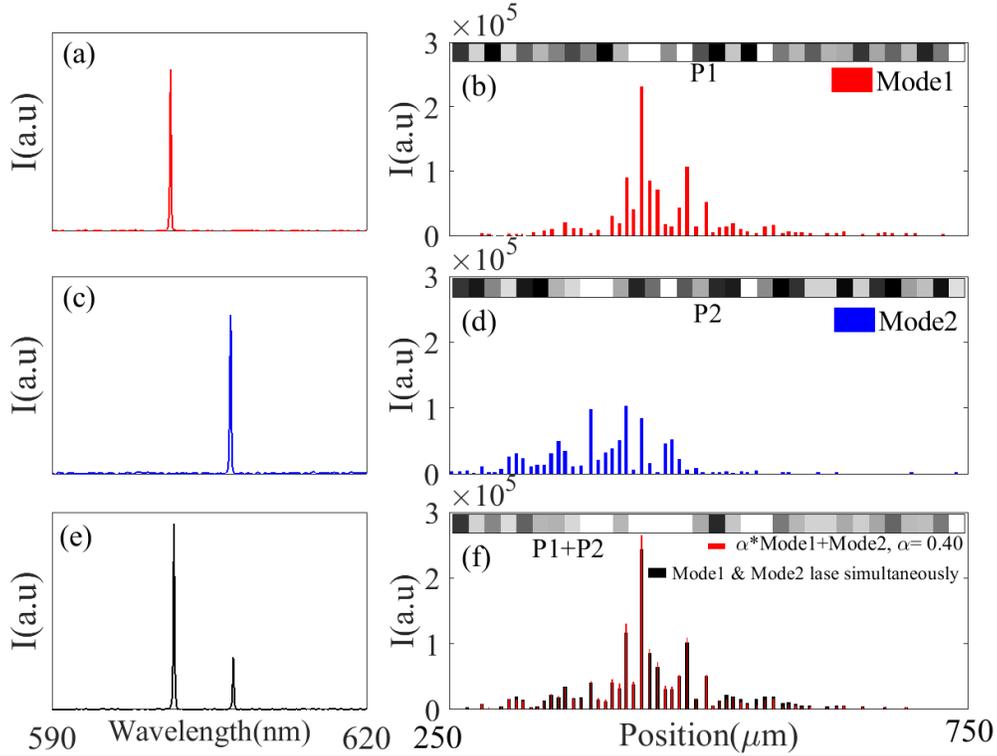

Fig. 6. **Robustness to cross-gain mode competition.** a&b: Emission spectrum and spatial intensity profile of lasing mode @608.00 nm. c&d: Emission spectrum and spatial intensity profile of lasing mode @602.72 nm. e&f: (Black bars) Emission spectrum and spatial intensity profile when both mode 608.00 nm and 602.72 nm lase together. (Red bars) The weighed sum of the intensity profile of mode 1 and mode 2. The weighing coefficient $\alpha$ is 0.40. Corresponding pump intensity profiles are shown on top of each intensity profile.

## 3. Conclusion

We have shown that nonlinear gain, gain saturation and nonlinear modal interaction do not modify the Anderson-like localized states. This is based on the preliminary demonstration that the modes of the passive systems provide with the necessary feedback for laser oscillations and are identical to the corresponding lasing modes away from the sample edges. This brings clear evidence that localization remains unaffected by gain-type nonlinearities. Whether localization is preserved, enhanced or hampered by other types of nonlinearity remains a matter of debate. There have been several theoretical [44–46] and experimental [47] investigating this fundamental question. Transverse localization of light has been tested in the presence of optical Kerr effect [48–51]. Other types of nonlinearity and their impact on localization have been investigated, such as thermal-induced and non-local nonlinearities [52] or quadratic nonlinearities [53–57]. Our method of investigation can be naturally extended to investigate other types of nonlinearity, since we can fully monitor the spatial profile of the localized modes individually. For instance, a material with large Kerr effect such as nematic liquid crystal [58], can be easily introduced in our random laser and its nonlinear effect on the localized modes can be tested.

That the mode profile remains unaffected under strong modal crosstalk and gain competition is probably the most striking result of our study. Actually, the robustness of the nonlinearly interacting localized modes offers the unexpected possibility to observe true interaction-induced mode switching (IMS) [43, 59]. In this steady-state effect, either of two gain-competing lasing modes with well-defined wavelength and well-preserved spatial profile is switched-off by the other's onset. It was argued in [43] that IMS was not expected to occur in random lasers because lasing mode pattern change dramatically above threshold and large cross-interaction is supposedly unlikely. Actually, this is not true in our case where modes are localized albeit significant spatial overlap, which results in robustness and possible strong cross-gain saturation. Complete IMS in our random laser is illustrated in the Supplement 1(section A). We believe our findings open interesting perspective to control mode switching in a disorder photonic system and can find potential applications in all-optical flip-flop memories [60], tunable sensitive switches [61], or highly sensitive sensors [62, 63].

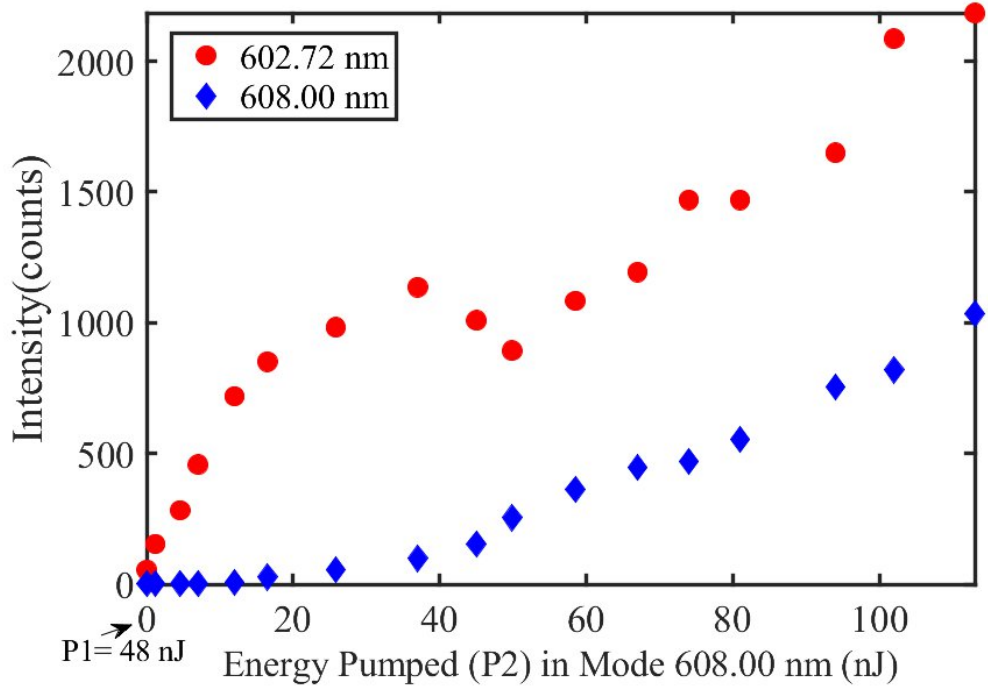

Fig. 7. **Mode competition.** Peak intensity evolution of mode 1 (red) and mode 2 (blue) with increasing pump energy of profile P2 at a constant pump energy P1=48 nJ.

**Acknowledgments.** We thank Prof. V. Freilikher for useful discussions. We are grateful to Dr. Yossi Abulafia for his help in the fabrication process and the Bar-Ilan Institute of Nanotechnology & Advanced Materials for providing with fabrication facilities. This research was supported by the Israel Science Foundation (Grants No. 1871/15, 2074/15 and 2630/20) and the United States-Israel Binational Science Foundation NSF/BSF (Grant No. 2015694). B. K. thanks the PBC Post-Doctoral Fellowship Program from the Israeli Council for Higher Education. P. S. is thankful to the CNRS support under grant PICS-ALAMO.

**Data availability.** Data underlying the results presented in this paper are not publicly available at this time but may be obtained from the authors upon reasonable request.

**Disclosures.** The authors declare no conflicts of interest.

See Supplement 1 for supporting content.

# Impact of Non-Hermiticity and Nonlinear Interactions on Disordered-Induced Localized Modes

BHUPESH KUMAR,[1,3] AND PATRICK SEBBAH[1,2,*]

[1]*Department of Physics, The Jack and Pearl Resnick Institute for Advanced Technology, Bar-Ilan University, Ramat-Gan, 5290002 Israel*
[2]*Institut Langevin, ESPCI ParisTech CNRS UMR7587, 1 rue Jussieu, 75238 Paris Cedex 05, France*
[3]*bhupeshg2@gmail.com*
[*]*patrick.sebbah@biu.ac.il*

**This document provide supplementary information to " Impact of Non-Hermiticity and Nonlinear Interactions on Disordered-Induced Localized Modes"**

### A. Cross-gain saturation induced mode switching in Random Laser.

Recently, the phenomenon of interaction-induced mode-switching (IMS) has been predicted for any standard lasing systems such as microdisk lasers and coupled one-dimensional optical cavities [1]. This mode-switching phenomenon is described as the simultaneous onset of a new lasing mode and turn-off of a previously lasing mode and is explained in terms of strong cross-gain saturation between the two modes [2]. Spatial intensity profile and lasing frequency of the competing modes remain unchanged. Mode-switching in optics has raised strong interest in the last two decades due to its potential application to all-optical flip-flop memories [3] and tunable sensitive switches [4, 5]. Mode switching has been observed experimentally in various photonic systems, including ring lasers [6], quantum dot micro-pillar lasers [7], perviskite [8] and plasmonics [9] microlasers, as wells as whispering gallery mode silica Raman laser [10]. Here, we experimentally demonstrate IMS of two interacting lasing modes in a random laser. In this experiment, a lasing mode (Mode1) at $\lambda = 606.33$ nm is selected with the optimized pump profile shown in Fig.S1(a). We then monitor the emission spectrum of the laser when the intensity of half the pump profile (P1) is progressively increased, while the other half (P2) is maintained (Fig. S1(c)). Starting from a singlemode spectrum at 606.33 nm, we observe that Mode1 decreases while a new lasing mode (Mode2) at $\lambda = 620.00$ nm starts lasing. Eventually, Mode1 is switched-off. Mode2, which has been turned on, is identified as the lasing mode selected by the half pump profile P2 alone (Fig. S1b). The process is reversible: Logarithm of peak intensity of mode1 (red) and mode2 (blue) are recorded and plotted as the pump energy in second half of the pump is increased again (Fig.S1(d)).

The onset of a new lasing mode at the expense of another lasing mode can be explained in terms of cross-gain saturation [2]. Strong competition for gain increases with spatial overlap of the two modes. By imaging individually the intensity spatial profile of the two lasing modes (Fig. S1(e)), we confirm that modal spatial overlap is significant, which explains why IMS is easily observed in our case. To the best of our knowledge, lasing mode switching induced by modal interaction has never been demonstrated in random lasers.

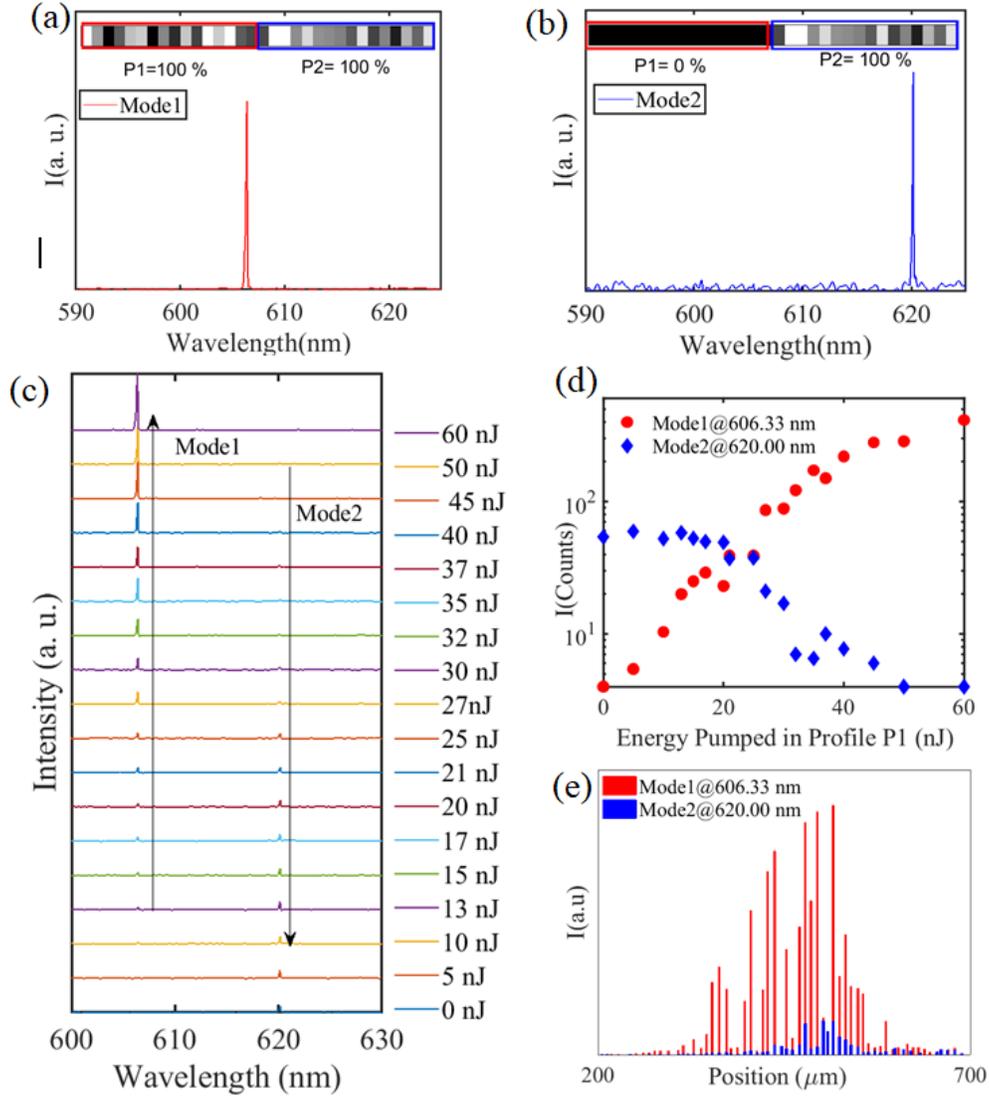

Fig. S1. **Mode switching in Random laser:** (a) Emission spectrum and optimized pump profile (inset) of target mode. (b) Recorded emission spectrum after assigning zero gray scale level to Pump profile P1. Modified profile shown in inset. (e) Emission spectrum recorded at various energy level of pump P1 and P2. (d) Peak intensity (log scale) of mode 1 and mode 2 as a function of energy pumped in profile P1. (e) Spatial Profile of Mode1 and Mode2.